\documentclass[aps,twocolumn,groupedaddress,showpacs,amssymb,prl,floatfix,preprintnumbers]{revtex4}
\usepackage{graphicx,color}

\newcommand{\slrr}      {$T_1^{-1}$}
\newcommand{\srcuo}    {SrCuO$_2$}
\newcommand{\srcacuo}   {Sr$_{0.9}$Ca$_{0.1}$CuO$_2$}
\newcommand{\srzweicuodrei} {Sr$_2$CuO$_3$}

\begin{document}

\thispagestyle{myheadings}

\title{Spin Gap in the Zigzag Spin-1/2 Chain Cuprate Sr$_{0.9}$Ca$_{0.1}$CuO$_2$ }

\author{F. Hammerath, S. Nishimoto, H.-J. Grafe, A.U.B. Wolter, V. Kataev, P. Ribeiro, C. Hess, S.-L. Drechsler, B. B\"uchner}

\affiliation{Leibniz Institute for Solid State and Materials Research IFW Dresden, P.O. Box 270116, D-01171 Dresden, Germany}

\date{\today}

\begin{abstract}

We report a comparative study of $^{63}$Cu Nuclear Magnetic Resonance spin lattice relaxation rates, \slrr, on undoped SrCuO$_2$ and Ca-doped \srcacuo\ spin chain compounds. A temperature independent \slrr\ is observed for \srcuo\ as expected for an $S=1/2$ Heisenberg chain. Surprisingly, we observe an exponential decrease of $T_1^{-1}$ for $T < 90$\,K in the Ca-doped sample evidencing the opening of a spin gap. The data analysis within the $J_1$-$J_2$ Heisenberg model employing density-matrix renormalization group calculations suggests an impurity driven small alternation of the $J_2$-exchange coupling as a possible cause of the spin gap.
\end{abstract}


\maketitle

\begin{figure}
\begin{center}
 \includegraphics[width= 6.5cm,clip]{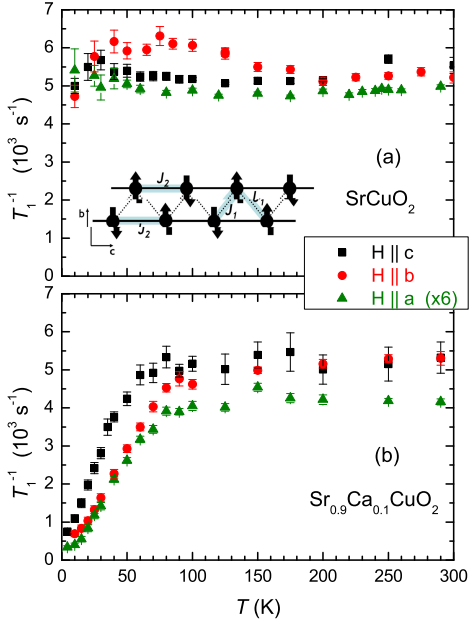}
 \caption{(color online) Spin lattice relaxation rate $T_1^{-1}$ of \srcuo\ (a) and \srcacuo\ (b) for different field directions. Inset: zigzag structure and  main couplings.}
 \label{T1SrCuO2}
\end{center}
\end{figure}

One-dimensional (1D) quantum magnets are intriguing model systems for studying correlated many-body quantum physics. They are theoretically well treatable which often yields clear-cut predictions about their ground states and excitation spectra (see, e.g., \cite{quantmag}). A particularly important case is the exactly solvable uniform antiferromagnetic (AFM) $S=1/2$ Heisenberg chain model (HM) which exhibits a \textit{gapless} spectrum of elementary spinon excitations and a ground state lacking long-range magnetic order. 
Spin gaps in the excitation spectra of closely related quantum systems arise only under certain circumstances. For instance they occur naturally in systems with a pronounced chemically caused dimerized structure or in spin-Peierls compounds with a spontaneously dimerized ground state \cite{Bray, Buzdin}. The latter become highly nontrivial in the presence of frustrated exchange interactions, being therefore a hot topic in the field of quantum magnetism. The parent compound \srcuo\ considered here, where Cu$^{2+}$ (3$d^9$, $S = 1/2$) ions bridged via oxygen ligands form well defined chain units along the crystallographic $c$-direction (see inset of Fig.~\ref{T1SrCuO2}), however, does not belong to either of them. It is commonly considered as one of the best realizations of the 1D $S=1/2$ AFM HM. Due to a very small and frustrated coupling between different zigzag chains it shows a N\'eel-like ordering below $T_N \approx 2$\,K \cite{MatsudaPRB1997} only, despite the huge effective AFM next nearest neighbor (NNN) exchange integral $J_2 \sim $ 2000\,K \cite{MotoyamaPRL1996, ZaliznyakPRL2004, KoitzschPRB2006}. The nearest neighbor (NN) coupling $J_1$ is much weaker, ferromagnetic (FM) and frustrated with an estimated ratio of $|J_1/J_2| \lesssim 0.1-0.2$ \cite{Rice1993}. Inelastic neutron scattering reports the onset of a weak sublattice magnetization below $T_{c1} = 5$\,K, static on a ns time scale, but the absence of 3D longe range order for $T \gtrsim 10^{-4} J/k_B$. Furthermore, it suggests a gapless spinon excitation spectrum for this compound down to the lowest measured energy transfer of 0.5\,meV \cite{ZaliznyakPRL1999}. However, despite inevitably present spin-phonon coupling, a spin-Peierls transition has not been found at low temperature, in contrast to the well-known CuGeO$_3$ \cite{Hase}. 

In this Letter we address this dichotomy by studying the effect of isovalent Ca substitution on the Sr sites in \srcuo, outside the spin chain unit. Our comparative Nuclear Magnetic Resonance (NMR) study investigates the temperature dependence of the $^{63}$Cu nuclear spin lattice relaxation rate \slrr\ for single crystals of \srcuo\ and \srcacuo. While $T_1^{-1}$ is temperature independent in \srcuo\ in compliance with the $S$=1/2 HM and the observed gapless spinon continuum \cite{ZaliznyakPRL1999}, an exponential decrease of $T_1^{-1}$ below 90\,K for \srcacuo\ evidences a significant gap in the spin excitation spectrum. We present a theoretical analysis of this striking observation in terms of the $J_1$-$J_2$ 1D HM for two possible scenarios: {\it (i)} the AFM-AFM frustrated HM for which, based on qualitative arguments only, a small spin gap has been predicted in the crossover region between a commensurate dimerized-like and an incommensurate spiral-like state \cite{White-Affleck96}; {\it (ii)} the standard spin-Peierls scenario with an alternation of the largest AFM NNN coupling $J_2$ in the presence of a realistic relatively small NN coupling $J_1$ of arbitrary sign. We conclude on the relevance of scenario {\it (ii)} for the occurrence of a spin gap in \srcacuo\ where subtle structural distortions due to Ca substitution might cause a small modulation of $J_2$ and trigger a drastic change of the spin excitation spectrum.

Single crystals of \srcuo\ and \srcacuo\ were grown by the travelling solvent floating zone method. Their high quality, i.e. single crystallinity, stoichiometry, the absence of phase irregularities, and second phase inclusions, was verified by x-ray diffraction, polarized optical microscopy, energy-dispersive x-ray spectroscopy and inductively coupled plasma optical emission spectrometry \cite{Ribeiro}. The Ca-content of \srcacuo\ was found to be (10.0 $\pm$ 0.2)\,\%. 
Structural refinement on \srcacuo\ showed no measurable Ca occupancy of the Cu sites. The averaged change of the Cu-O-Cu bond angle and bond distance along the direction of $J_2$ is only 0.35\,\%. The $^{63}$Cu NMR spin lattice relaxation rate, $T_1^{-1}$, was measured by inversion recovery in a magnetic field of $H = 7.0494$\,T for $H\parallel a$, $b$ and $c$, respectively. The values of \slrr\ were obtained by fitting the recovery curves to the standard expression for magnetic relaxation of a nuclear spin $I=3/2$ \cite{McDowell,TakigawaPRB1998} yielding one single $T_1$ component at any temperature \cite{comment0}.

For \srcuo, $T_1^{-1}$ remains constant over the whole temperature range, and increases only slightly at low temperature [Fig.~\ref{T1SrCuO2}(a)] \cite{comment1}. Note that the error bars increase with decreasing temperature due to a strong broadening of the NMR resonance lines at low temperature \cite{else}. The observed behavior strongly supports the gapless excitation spectrum in \srcuo\ and is consistent with theoretical calculations for the $S$=1/2 AFM HM predicting 
a temperature independent $T_1^{-1}$ \cite{TakigawaPRL1996, Sachdev, Sandvik}.
\begin{figure}
\begin{center}
 \includegraphics[width=\columnwidth,clip]{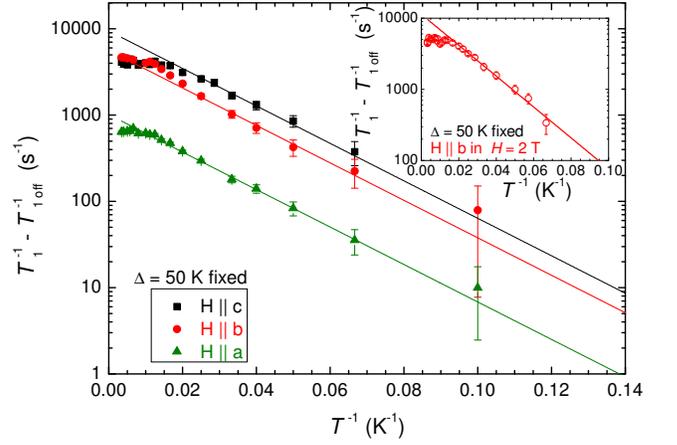}
 \caption{(Color online) Spin lattice relaxation rate $T_1^{-1}$ of \srcacuo\ vs. inverse
 temperature. Lines are fits of Eq.~(\ref{T1}) with a fixed spin gap of $\Delta = 50$\,K at low temperature (10 - 40\,K).  For clearness $T_{1,\mbox{\tiny off}}^{-1}$ has been subtracted from the data and the fits. Inset: the same for $H \| b$  in $H$ = 2\,T.}
\label{delta50K}
\end{center}
\end{figure}
Fig.~\ref{T1SrCuO2}(b) shows the temperature dependence of $T_1^{-1}$ for \srcacuo. Similar to the undoped sample, $T_1^{-1}$ shows the expected temperature independent behavior for $T > 90$\,K, but unexpectedly decreases exponentially by more than one order of magnitude for $T < 90$\,K \cite{comment2}.

In case of magnetic relaxation the spin lattice relaxation rate probes directly the imaginary part of the dynamic susceptibility $\chi''$ of the electronic spin system:
\begin{equation}
T_1^{-1} \propto \frac{T}{const.} \sum_{\vec{q}} A_{\bot}^{2}(\vec{q},\omega) \cdot \frac{\chi''(\vec{q},\omega)}{\omega}.
\end{equation}
Here $A_{\bot}$ is the hyperfine coupling, $\vec{q}$ the wave vector and $\omega$ the NMR frequency. The observed exponential decrease of $T_1^{-1}$ gives therefore evidence for the opening of a gap $\Delta$ in the excitation spectrum of \srcacuo\ that is manifested by an exponential reduction of $\chi''$. We estimated the magnitude of the gap by fitting our data to an activated temperature dependence:
\begin{equation}
T_1^{-1}(T)= T_{1,\mbox{\tiny off}}^{-1} + const.\cdot e^{(-\Delta /T)} \label{T1},
\end{equation}
which applies for $T< \Delta $. To account for the non-vanishing value of $T_{1}^{-1}$ as $T\rightarrow 0$, we added a constant offset $T_{1,\mbox{\tiny off}}^{-1}$ to the activation behavior. The results shown in Fig.~\ref{delta50K} clearly demonstrate that for $T\leq40$\,K the decrease of $T_{1}^{-1}$ can be well described with an isotropic spin gap of $\Delta = 50$\,K. A possible field dependence 
could be excluded by supplementary measurements of $T_1^{-1}$ in $H =2$\,T, also yielding $\Delta = 50$\,K (see inset of Fig.~\ref{delta50K}).

The opening of a spin gap suggests that the magnetic properties of the Cu spin chains change drastically by substituting Sr with isovalent Ca. Note that neither in susceptibility nor in other thermodynamic properties any indication for a spin gap was found. This is not surprising since owing to the very large value of $J_2$ the intrinsic static magnetic susceptibility at $T\ll J_2$ is very small. Therefore the thermodynamics might just overlook the gap since the total static magnetic response at low $T$ is dominated by other contributions, such as the Van-Vleck susceptibility and Curie-like impurity contributions.

Two different approaches to explain such a doping-induced spin gap in \srcacuo\ will be discussed based on the following spin Hamiltonian written in the adiabatic approximation: 
\begin{eqnarray}
H = J_1 \sum_i \vec{S}_i \cdot \vec{S}_{i+1} + J_2 \sum_i (1+(-1)^\alpha \delta) \vec{S}_i \cdot \vec{S}_{i+2}. \label{hamiltonian}
\end{eqnarray}
Here $\vec{S}_i$ is a spin-$\frac{1}{2}$ operator at site $i$ and $\alpha=\sqrt{2}\sin\frac{\pi}{4}(2i-1)$, $J_1$ and $J_2$ are the NN and the NNN exchange interactions and $\delta$ is the modulation of $J_2$. The spin gap is evaluated as the energy difference between the lowest triplet state and the singlet ground state,
\begin{equation}
\Delta(L)=E_1(L)-E_0(L), \ \ \ \Delta=\lim_{L \to \infty}\Delta(L),
\end{equation}
where $E_n(L)$ is the $n$-th eigenenergy ($n=0$ denotes the ground state) of the system with length $L$. We employ the density-matrix renormalization group (DMRG) technique, a powerful numerical method for various 1D quantum systems \cite{White92}. Open-end boundary conditions are applied in the chain direction and it enables us to calculate the energies of the ground-state and low-lying excited-states quite accurately for very large but always finite systems (up to the order of $\sim 10^3$ sites). 

\begin{figure}[]
    \includegraphics[width= 5.5cm,clip]{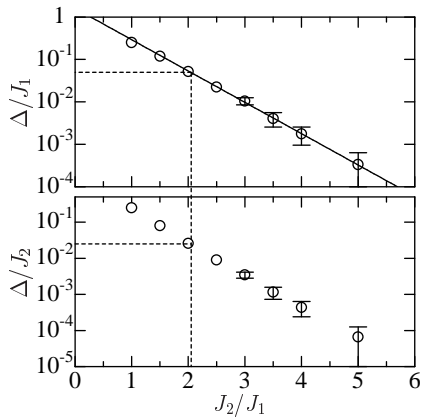}
  \caption{Spin gap as a function of $J_2/J_1$ in the AFM-AFM sector of the $J_1$-$J_2$ HM. The experimental value of the spin gap in units of $J_1$ is depicted by dashed lines adopting $J_2=$\,2000\,K.}
\label{Theo1}
\end{figure}

{\em Scenario (i)} gives attention to the frustrated NN coupling $J_1$. For FM NN interaction $J_1<0$, whose absolute value is much smaller than the NNN interaction $J_2$, as it is likely the case in \srcuo, the zigzag chain is not expected to show a spin gap \cite{White-Affleck96}. However, in the case of a small AFM NN coupling $J_1>0$, the zigzag chain should exhibit a spin gap which scales exponentially with the ratio of the two competing couplings \cite{White-Affleck96}:
\begin{equation}
\Delta \propto \exp{\left(-const.\cdot \frac{J_2}{J_1}\right).} \label{gap}
\end{equation}
According to the Goodenough-Kanamori-Anderson (GKA) rules \cite{AndersonPR1959} a Cu-O-Cu bond angle of $180^\circ$ yields a strong AFM exchange between the Cu spins ($J_2$), while an angle of $90^\circ$ results in a weak FM coupling ($J_1$). However, as soon as the bond angle slightly deviates from $90^\circ$, the interaction $J_1$ can become AFM again \cite{braden}. Local structural distortions in \srcacuo\ induced by Ca ions could yield such a deviation, resulting in a local spin gap corresponding to Eq.~(\ref{gap}).

Within this scenario ($\delta=0$ and  $J_1 > 0$) we study chains with several lengths $L = 512$ to
$4096$, keeping $m=1200$ density-matrix eigenstates in the renormalization procedure. In this way, the discarded weight, $w_{\rm d}$, is less than $10^{-7}$, while the maximum error $\Delta E$ in the ground-state and low-lying excited states energies is less than $10^{-7}-10^{-6}$. The extrapolated results of the spin gap are shown in Fig.~\ref{Theo1}. For $J_2/J_1 \ge 2$, the spin gap can be well fitted by the expression
\begin{equation}
\Delta/J_1=1.6\exp \left( -1.7J_2/J_1 \right),
\end{equation}
which might be useful for the description of frustrated systems in general. However, for the zigzag chain under consideration with the experimentally obtained gap of $\Delta = 50$\,K a value of $J_1 \sim 1000$\,K would be required to explain our NMR data. This is unrealistically large for a Cu-O bond angle only slightly below or above 90$^\circ$ \cite{braden, comment3}. Hence, scenario \textit{(i)} can be discarded.

{\em Scenario (ii)} examines the possibility of an alternating AFM NNN exchange $J_2(1\pm \delta)$, which is known to cause an intrinsic spin gap. Within this scenario we assume that the Ca substitution might induce either directly or via a softening of the lattice and a concomitant strongly increased spin-phonon interaction a modulation of the NNN coupling constant $J_2$, whose simplest realization is a periodic alternation. The validity of this scenario will be examined for FM and AFM $J_1$.

\begin{figure}[]
    \includegraphics[width= \columnwidth,clip]{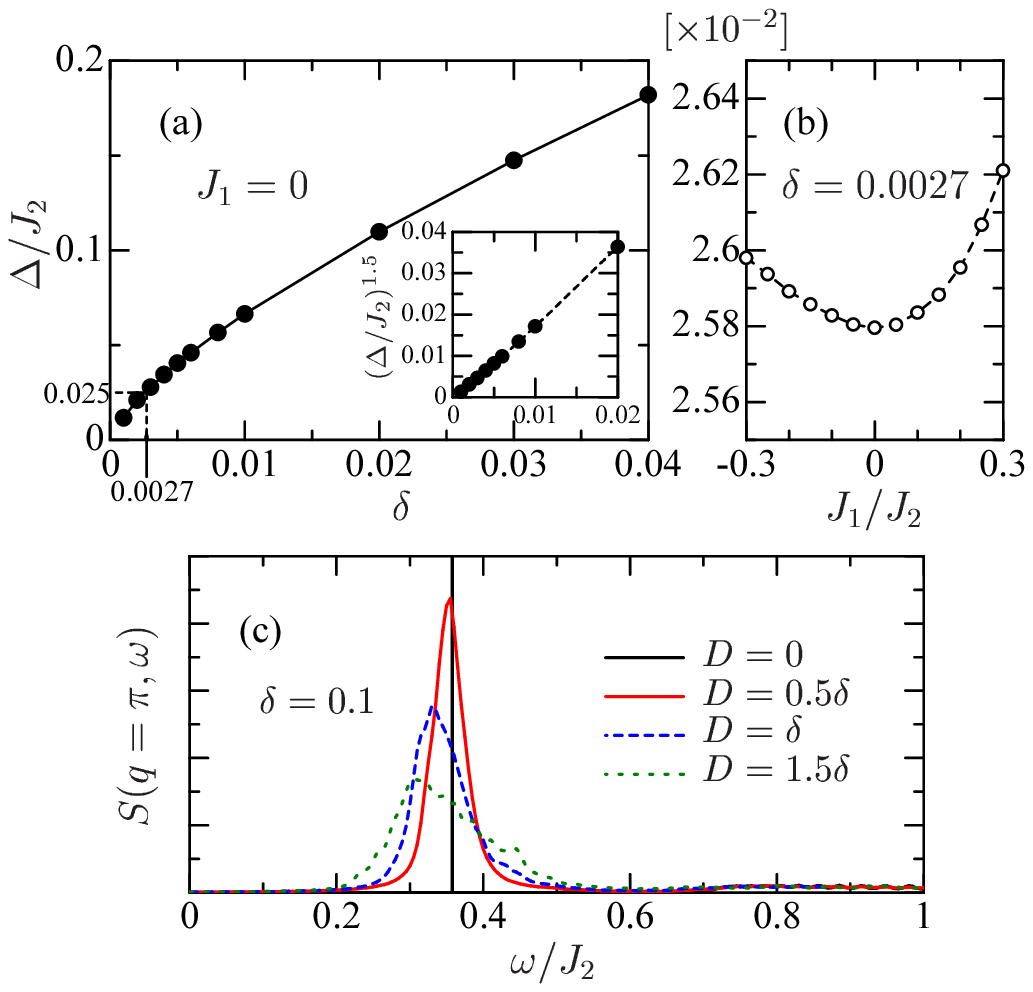}
  \caption{The same as in Fig.~\ref{Theo1} for an alternating AFM NNN exchange $J_2(1 \pm \delta )$ at (a) zero and (b) weak arbitrary NN-exchange $J_1$. (c) Spin structure factor for several disorder strengths $D$ with $\delta=0.1$ and $J_1=0$. The inset of (a) shows $({\Delta}/{J_2})^{1.5}$ for small alterations $\delta$.}
\label{sc2}
\end{figure}

We study chains with several lengths $L = 64$ to $392$, keeping $m=2000$, which ensures $w_{\rm d} < 10^{-14}$ and $\Delta E < 10^{-12}$. In Fig.~\ref{sc2}(a), the spin gap is plotted as a function of $\delta$ for $J_1$ = 0. We estimate $\delta=0.0027$ to reproduce the experimental value of the spin gap $\Delta=50$\,K for $J_2=2000$\,K. Next, we introduce the NN exchange interaction $J_1$. The spin gap is calculated as a function of $J_1$ with fixed $\delta=0.0027$ [Fig.~\ref{sc2}(b)]. We find that the value of $\Delta$ is hardly affected by $J_1$. Thus, only a small alternation factor $\delta=0.0027$ is needed to reproduce the spin gap found by NMR, irrespective of the strength and sign of the interchain coupling $\left| J_1 	\right| \ll J_2 $.

Beyond the simple alternation of $J_2$, a more realistic situation would be given by including also the direct impact of disorder in the exchange coupling caused by the Ca impurities. Thus, we consider an additional term to Eq. (\ref{hamiltonian}), $H_{\rm dis}=2D \sum_{i=1}^{L-2} \varepsilon_i \vec{S}_i \cdot \vec{S}_{i+2}$, where $\varepsilon_i$ is defined by a box probability distribution ${\cal P}(\varepsilon_i)=\theta(1/2-|\varepsilon_i|)$ with the step function $\theta(x)$ and the disorder strength is controlled by $D$. We calculate the spin structure factor $S(q=\pi,\omega)$ on random sampling $30$ realizations of ${\cal P}(\varepsilon_i)$ for $L=384$, using dynamical DMRG \cite{Jeckelmann2002}, and take an average of the results for each $D$. Fig.~\ref{sc2}(c) shows the averaged $S(q=\pi,\omega)$ for several $D$ values with $\delta=0.1$ and $J_1=0$. We find that the delta peak for $D=0$ is broadened by the disorder and a "pseudo spin gap"-like behavior appears. Note that a relatively large $\delta=0.1$ is chosen to be able to clearly follow the $D$-dependence. For $D>\delta$ the singlet pairs are partially collapsed, i.e., magnetic moments are locally induced, and the spectral weight at $\omega=0$ becomes finite. This must be related to the occurrence of a finite, doping-dependent residual relaxation rate $T_{1,\mbox{\tiny off}}^{-1}$ [see Eq. (\ref{T1}) and Fig.~\ref{T1SrCuO2}(b)].

In summary, measurements of the $^{63}$Cu NMR spin lattice relaxation rate $T_1^{-1}$ in a single crystal of the $S$ = 1/2 chain compound \srcacuo\ reveal a striking exponential decay of $T_1^{-1}$ below 90\,K in sharp contrast to a constant $T_1^{-1}$ in the parent compound \srcuo\ \cite{comment}. The observed decay points to the opening of a spin gap with $\Delta=50$\,K. Our DMRG calculations show that this spin gap can be well reproduced within the $J_1$-$J_2$ 1D HM considering an alternating AFM NNN $J_2(1 \pm \delta )$ with only a small alternation of about $\delta=0.0027$, regardless of the strength and sign of the NN $J_1$. Such a remarkable sensitivity of the low energy excitation spectrum of the $S$ = 1/2 chains in \srcuo\ to structural substitutional defects outside the zigzag chain appears to be crucial for the understanding of the spin dynamics in low dimensional cuprates. Our results call therefore for extensive experimental and theoretical work on this and related systems to understand the impact of additional degrees of freedom on the fundamental magnetic phenomena occurring in quantum spin magnets on the basis of cuprates. 
For example diffraction experiments may confirm the dimerization \cite{comment4}, while neutron scattering, resonant inelastic x-ray spectroscopy, optics and specific heat may observe the spin gap and/or a possible phonon softening. So far there seem to be no generally accepted microscopic models for spin-Peierls physics in cuprates. We hope that our work will stimulate the development of such models. 

We thank J. Sirker, M. Vojta and A. Zheludev for fruitful discussions. This work has been supported by the DFG through FOR 538 (Grant No. BU887/4), FOR 912 and through projects HE3439/7, HE3439/8, DR269/3-1 and WO 1532/3-1.



\begin{thebibliography}{99}
\bibitem{comment0} Due to an overlap of the $^{63}$Cu central line with a $^{65}$Cu satellite for $a$ and $c$, $T_1^{-1}$ was measured at the $^{63}$Cu satellite for these directions.
\bibitem{comment2} At low temperature a frequency dependence of $T_1^{-1}$ was found along the broad resonance lines for \srcacuo. The data in Fig.~\ref{T1SrCuO2}(b) show $T_1^{-1}$ measured at the center of the resonance lines. Away from the center $T_1^{-1}$ is slightly enhanced, but still remarkably reduced compared to the values for $T>90$\,K.
\bibitem{comment} Note that a decreasing \slrr\ has been observed earlier in nominally undoped \srcuo\ powder samples \cite{IshidaPRB1996}. However, we can rule out the assumption that the FM frustrated $J_1$ causes the decrease of \slrr\ (i) by the \slrr\ data of our clean \srcuo\ single crystals as well as (ii) theoretically by showing that the observed spin gap is practically independent of the strength and the sign of $J_1$.
\bibitem{comment4} In cuprates the main exchange integral $J$ scales with the Cu-O-distance $r$ as $J \sim r^{-n}$, where $n = 5-7.7$ \cite{Kawada}. Adopting $\delta = 0.0027$, this results in 
structural changes of $(1-2)\cdot10^{-3}$\,\AA\ in the gapped phase of \srcacuo, which will be challenging to observe.
\end{thebibliography}
\end{document}